\documentclass[prd,aps,preprint,letterpaper,noeprint,longbibliography,nofootinbib, superscriptaddress]{revtex4-1} 

\usepackage[colorlinks, allcolors=blue]{hyperref}

\usepackage{amsmath}  
\usepackage{graphicx} 
\usepackage{comment}
\usepackage{tikz}

\usepackage[font=small,labelfont=bf,justification=justified,format=plain]{caption}

\usepackage{subcaption}
\usepackage{nicefrac}  
\usepackage{comment}
\usepackage[margin=0.9in]{geometry}
\newgeometry{includefoot,left=2cm,right=2cm,bottom=1cm,top=2cm}
\let\oldfootnote\footnote
\renewcommand{\footnote}[1]{%
    \begingroup%
    \linespread{1}
    \oldfootnote{#1}%
    \endgroup%
}

\begin{document}
\count\footins = 1000

\title{Self-Interacting Dark Matter and flavor anomalies in an atlas of leptophilic models}

\author{Nicolás Gómez} 
\email{nigomezcr@unal.edu.co}
\affiliation{Departamento de F\'isica. Universidad Nacional de Colombia, Bogotá D.C., Colombia.}

\author{Andrés Castillo}

\affiliation{Departamento de F\'isica, Universidad Antonio Nari\~no, Bogot\'a D.C., Colombia}

\begin{abstract}
We consider an atlas of leptophilic models with non-universal charges that combine the quantum numbers \((L_e - L_{\mu})\) and \((L_{\mu} - L_{\tau})\) in various realizations. We explore diverse manifestations of the atlas to understand their impact on explaining the muon \((g-2)_{\mu}\) anomaly, as well as the origin of self-interacting dark matter, which serves as a feasible explanation for the small-scale challenges of cold dark matter.  We found that a single vector gauge boson with mass range of $\sim 10~\mathrm{MeV}$ and non-universal couplings as indicated successfully mediates interactions in the fermion and dark matter sectors in such a way that $(g-2)_{\mu}$ additional contribution agrees with experimental constraints and the self-interacting dark matter fluid behaves as suggested by observations at different astrophysical length scales. Furthermore, the model requires a dark matter candidate within $10-100~ \mathrm{GeV}$ order mass and a strong dark coupling ($g_{\chi} \sim 0.3$, much stronger than the lepton's couplings $g'_l \lesssim 10^{-4}$) to be a successful scenario to unravel astrophysical small scale problems.  

\end{abstract}

\maketitle


\section{Introduction}

One of the remaining pieces in the puzzle of the Standard Model (SM) of particle physics is the nature of Dark Matter (DM). From its gravitational effects, we know that, at large scales, DM should be a cold (non-relativistic) and collisionless fluid (cold dark matter, or CDM). Although these features have been incorporated long ago in the $\Lambda$CDM cosmological model (where $\Lambda$ stands for the cosmological constant), they have been challenged by observations in recent years upon looking at small galactic halos~\cite{2015AJ....149..180O, 2020ApJS..247...31L}. The most popular of the so-called \textit{small scale challenges} is the core-cusp problem: a discrepancy between observed and only DM simulated density profiles of halos surrounding small DM dominated galaxies~\cite{2017ARA&A..55..343B}. The other small scale problems being the missing satellites (a discrepancy in the number of small satellite subhalos) and the too-big-to-fail (an absence of more luminous small galaxies inside DM subhalos)~\cite{2015AJ....149..180O}. Added to these, a \textit{diversity problem} has also been alluded. The diversity problem consists on a large diversity in the distribution of matter (as deduced by a diversity in shape of the velocity profiles) inside halos of the same mass, this variety is larger than what is predicted for a self-gravitating fluid with no interactions or dissipation, for which velocity profiles should be more uniform among halos of the same mass~\cite{Oman:2015xda, Creasey:2017qxc, KuziodeNaray:2009oon}. 

Recently, in particular, a set of gas-rich Ultra Diffuse Galaxies (UDGs) has been discovered, which have a deficit of mass inconsistent with the average halos of CDM simulation \cite{ManceraPina:2020ujo, Kong:2022oyk}. At the other extreme, another recent observation reported  that a highly dense substructure perturbs the strong lens SDSSJ0946+1006 in a manner that is difficult to explain with CDM~\cite{Minor:2020hic}. Both phenomena add up to a clear difficulty that faces the standard cosmological dark matter model in explaining small scale observations.

It seems that baryonic effects can not completely explain the discrepancy between only-DM simulation and observation at these scales alone~\cite{Nadler:2023nrd}. In principle, baryons expelled by the highly energetic supernova explosions redistribute DM itself, lowering density cusps and producing flat-shaped internal density distributions, which might solve the core-cusp problem~\cite{Navarro:1996bv}. Other environmental effects such as tidal stripping or ram pressure stripping due to a large galaxy, group of galaxies or cluster, impact strongly on the density distribution of its satellites, like for some UDGs. The primary role of cosmic reionization can also be relevant during the star formation period of the galaxies, which also impacts their baryonic matter content and dark matter distribution~\cite{Sales:2022ich}. However, these explanations are not enough to support $\Lambda$CDM against the increasing amount of discrepancies altogether~\cite{Tollet:2015gqa,DiCintio:2013qxa,Chan:2015tna, Fitts:2016usl, Pawlowski:2015qta}.  For example, it does not correctly explain the dark matter distribution inside the group of UDGs from Ref. \cite{ManceraPina:2020ujo}, and the later two discrepancies cannot be simultaneously explained by collisionless DM particles plus stellar processes, since the two sit at opposite extremes with respect to CDM predictions: UDGs seem to have far less dark matter than predicted, while the substructure is anomalously denser than expected.

 An alternative scenario with particular interest to both astrophysics and particle physics communities is dark matter with strong self interactions that produce a large scattering cross section, known as self-interacting dark matter (SIDM)~\cite{Spergel:1999mh}.  The self-, non-gravitational interactions of the particles composing dark matter fluid explain the formation of cores in small scale subhalos, and the diversity of velocity dispersion shapes at fixed mass, according to simulations~\cite{Tulin:2017ara}. The physics behind this is in the formation history of these halos: First, a dark matter halo with a Navarro-Frenk-White (NFW) density profile\footnote{The NFW density profile is a phenomenological model -based on large-scale cosmological simulations using the CDM model- that describes the distribution of dark matter within galaxies and galaxy clusters~\cite{1996ApJ...462..563N}.} is hotter in the outer part of the halo, the self-interactions then allow the efficient transfer of heat towards the interior. In doing so, the central density decreases, the collisions between DM particles happening several times per Hubble time settle the  thermal equilibrium~\cite{Moore:1999gc}, and a central density core is formed, with all particles in that region reaching thermal equilibrium. Afterward, matter is again pulled towards the center; density increases and heat is distributed back to the external part of the halo. This process is called gravothermal core-collapse, and it seems to adequately reproduce the observations~\cite{Zavala:2012us, Elbert:2014bma, Sagunski:2020spe, Correa:2020qam}.

Besides to its small-scale ($\lesssim1~ \mathrm{Mpc}  $) phenomenology being consistent with observations, the large-scale  behavior of SIDM models needs to match CDM successful predictions.  For the self-interaction cross section, this implies that it must depend on the velocity, such that low-relative-velocity particles have higher cross section than fast-moving ones. SIDM with this velocity dependence is commonly referred to as a velocity-dependent SIDM (vdSIM). Therefore, at small scales, vdSIDM particles might form cores after a gravothermal core-collapse evolution~\cite{Zavala:2012us, Elbert:2014bma, Sagunski:2020spe, Correa:2020qam} and at large scales it remains consistent with all successes of CDM.
A great variety of model independent analyses have been elaborated within this simple framework~\cite{Patel:2022qvv, Colquhoun:2020adl, Kaplinghat:2015aga}. On the other hand, from the point of view of particle physics, it has been shown that the simple scenario of a light (around MeV order) scalar or vector boson mediating self-interactions can produce the desired velocity dependence in the cross section~\cite{Spergel:1999mh,Tulin:2013teo}. 

The nature of that mediator tackling the CDM challenges is also a subject of  Beyond Standard Model (BSM) physics~\cite{Kamada:2018zxi, Heeck:2022znj}. Processes involving new interactions are also determined by a coupling constant associated to a new symmetry  and a new conserved charge, as well as the masses of the fermions and the mediator itself. The SM particles can be charged by this new dark force (a new gauge group). If such is the case, then: 1) a non-gravitational window is open for detecting dark matter, and 2) it must have been produced thermally, impacting the relic density that can be compared with the measurements done by Planck Satellite~\cite{2020A&A...641A...6P}. 

Currently, particle physics has discrepancies among theory, experiment and simulation on its own. A few sets of anomalous experimental results are still to reach a firm $5\sigma$ status to claim discovery of New Physics (NP): cosmic rays excesses~\cite{Chang:2018bpt, Bulbul:2014sua, HAWC:2018szf, Bartlett:2022ztj, Arcadi:2018pfo, Leane:2018kjk}, the $W$ mass anomaly~\cite{Borah:2022obi, Batra:2022pej, Borah:2022zim, Batra:2022org, Dcruz:2022dao, Chowdhury:2022jde}, or the muon anomalous magnetic moment $(g-2)_{\mu}$  discrepancy~\cite{Muong-2:2006rrc, Muong-2:2021ojo, Aoyama:2020ynm, Heeck:2022znj, Chowdhury:2022jde}, only to name a few. 

The effects of dark matter with non-gravitational interactions and their impact in galactic (sub)halos and anomalies in the particle physics experiments can be connected. If physics from BSM is  the cause of these anomalies, a plethora of particle physics models are available as long as they confront a set of well stablished experimental constraints. Among them, however, a very well motivated class of leptophilic models are $L_i - L_j$ models, where $L$ refers to lepton number and $i, j=e, \mu, \tau$ to the family. In particular, $L_{\mu} - L_{\tau}$ model is popular to address questions like the muon anomalous magnetic moment or the origin of neutrino masses ~\cite{Heeck:2022znj, Kamada:2018zxi}. 

A departure from the minimal scenario of $L_{\mu}-L_{\tau}$ consisting in a family or \textit{atlas} of minimal $U(1)'$ models, is also worth contemplating. Motivated by the atlas of leptophilic $U(1)_l$ models from Ref. ~\cite{Ballett:2019xoj} (subset of the complete anomaly free atlas described in Ref.~\cite{Allanach:2018vjg}), we consider a family of $U(1)_l$ models where the mediator is a light neutral $Z'$ vector boson (index $l$ stands for a leptonic symmetry). The gauge boson couples to leptons with flavor dependent charges ${Q'}_{i=e, \mu, \tau}$, which we take it depending on two parameters: $\rho$ and $\vartheta$, that are also used to label the models. In a pragmatic sense, this model is inspired by the future and current experimental possibilities in the leptonic sector, allowed by neutrino laboratories such as DUNE~\cite{DUNE:2020lwj, DUNE:2020ypp}, Miner$\nu$a \cite{MINERvA:2013zvz}, T2K and Super-Kamiokande \cite{T2K:2024wfn}, or lepton accelerators like BESIII \cite{BESIII:2009fln}, since any flavor violating effects could finally be manifest, as detailed in~\cite{Ballett:2019xoj}. At the same time, new physics effects could be responsible for the discrepancy in the muon's anomalous magnetic moment, currently  at the $5.1\sigma$ level~\cite{Muong-2:2006rrc, Muong-2:2021ojo, Muong-2:2023cdq}, with still some discussions on the role of the Hadronic Vacuum Polarization contribution to be done~\cite{Borsanyi:2020mff, Ce:2022kxy, ExtendedTwistedMass:2022jpw, Benton:2023dci}.

Similarly, in recent years the electron's anomalous magnetic moment has been measured with a $2.4\sigma$ significance with the Standard Model prediction, hinting towards New Physics~\cite{Davoudiasl:2018fbb, Parker:2018vye}. For tau lepton, in turn, new proposals regarding the measurements haven been suggested for future experimental searches, including the LHC~\cite{Beresford:2024dsc}. Hence, a full analysis of $(g-2)_{e, \mu, \tau}$ with the $Z'$ coupled to all lepton families seems relevant ahead  of the near future test of flavor violation in the leptonic sector. 


Previously, some works showed that $U(1)'$ gauge extensions that couple DM to the SM always require couplings to leptons, unless additional fermions that transform non-trivially under the SM gauge group are introduced~\cite{Ellis:2017tkh}. Seemingly, our model explores flavor violation in the leptonic sector with a chiral anomaly free atlas of leptophilic $U(1)$ models introducing at the same time self interacting DM candidate.

Likewise, if a dark matter fermion candidate is incorporated, which is charged under the new symmetry $U(1)$, self-interactions presented in the new dark sector might lead to the core-formation during halo formation since it behaves as SIDM candidate with a mediator exchange. The large scale behavior resembles that of collisionless CDM due to the velocity dependence of the cross section. We include such a new fermion, and describe its interaction and annihilation properties; confronting dark matter relic density and astrophysical effects in small structure formation. 

For the reasons outlined above, we consider such a leptophilic extended model and explore its impact on the anomalous magnetic moment of the leptons $(g-2)_{l}$, dark matter thermalization in the early universe, and the self-interacting capabilities of a plausible dark matter candidate in the astrophysical context of small-scale challenges. We highlight some novel features of the present work, which include the extension of the $L_{\mu} - L_{\tau}$ model to a family of models with non-universal couplings to the three generations in the context of the leptons' anomalous magnetic moments. In a similar manner, we provide, from a particle physics standpoint, a comparison between two particular cross sections with thermal effects in halo formation; commonly discussed in astrophysics literature~\cite{Yang:2022hkm}. Moreover, the small scale problems and the potential solution with a SIDM candidate, both extensively treated, are complemented with the new observations from UDGs and DM-dominated galaxies in the present analysis.

Our paper is organized as follows: In Sec.~\ref{section: Atlas of Models}, we introduce the atlas of leptophilic models with a DM candidate and a mediator $Z'$, and the charges of the particles in different benchmark chiral anomaly-free scenarios. Then we use them to explore the $Z'$ contribution to the anomalous magnetic moment of the muon in Sec.~\ref{section: Muon g-2}. Next, we discuss the dark matter cosmology of our candidate in Sec.~\ref{section: Production of Dark Matter Particles} and its role inside small subhalos in Sec.~\ref{section: Small Scale Effects}. Finally, we conclude and provide some remarks in Sec.~\ref{section:Conclusions}.

\section{Atlas of Models}\label{section: Atlas of Models}

The atlas of models consists on a family of $U(1)'$ extensions to the SM gauge group ($SU(2)_L\times U(1)_Y$) with a $Z'$ boson always coupled to the three families of leptons in a non-universal manner (i.e. with charges depending on the flavor). For a leptophilic $U(1)'$ group, the following configurations are the emergent chiral anomalies: 
\begin{subequations}
    \begin{align}
    &SU(2)_L \times SU(2)_L\times U(1)', \\
    &U(1)_Y \times U(1)_Y \times U(1)', \\
    &U(1)_Y \times U(1)' \times U(1)', \\
    &U(1)' \times U(1)' \times U(1)', 
\end{align}
\end{subequations}
whose cancellation is guaranteed by the following conditions on the charges respective to the gauge group

\begin{subequations}
    \begin{align}
     &  \sum_i^3 Q'_{L_i} = 0, \label{Chiral anomalies a}\\
     &  \sum_i^3 (3Q'_{L_i} - 6 {Q'}_{e_i}) = 0,  \label{Chiral anomalies b}\\
     &  \sum_i^3 ( {Q'}_{L_i}^2 -  {Q'}_{e_i}^2)=0,  \label{Chiral anomalies c}\\
     &  \sum_i^3 ( 2 {Q'}_{L_i}^3 - {Q'}_{e_i}^3 - {Q'}_{N_i}^3 )=0, \label{Chiral anomalies d}
    \end{align}
\end{subequations}
besides,  the gauge gravity anomaly is present as well, given by
\begin{equation}
    \sum_i^3 (2Q'_{L_i} - Q'_{e_i}- Q'_{N_i})=0;
    \label{Gravitational anomaly}
\end{equation}
where $Q'$ refers either to the value of the new charge for left-handed lepton doublet (with subscript $L_i$) or to the right-handed lepton singlet (with subscript $e_i$) for the $i$-th family. 
The charge for each lepton family is arbitrary, up to a free-anomaly assignment of two parameters, $\rho$ and $\vartheta$. We can see that if we parameterise the charge as

\begin{equation}
    Q' = \rho (L_e - L_{\mu}) + \vartheta (L_{\mu} - L_{\tau} ) ;\\
    \label{Atlas parameterisation}
\end{equation}
then, Eq.~\eqref{Chiral anomalies a} and \eqref{Chiral anomalies b} are trivially satisfied. On the other hand, if we take the value of the charges to be the same within family for neutrinos, left- and right-handed charged leptons (${Q'}_{L_i} = {Q'}_{e_i} = {Q'}_{\nu_i} = {Q'}_{i}$), then, Eq.~\eqref{Chiral anomalies c} is also satisfied. 
As can be seen on Eqs.~\eqref{Chiral anomalies d} and \eqref{Gravitational anomaly}, the model can accommodate additional right-handed neutrinos $N_i$. From the anomaly freedom requirement, the charges on those RH neutrinos should satisfy

\begin{equation*}
    \rho (\rho - \vartheta) = \sum_{i} {Q'}_{N_i}^3, 
\end{equation*}

where $Q'_{N_i}$ refers to the charge of the right-handed neutrinos. If we assign the same value of the charges within family, that is, if $Q'_{N_i} = Q'_i$ for $i = 1, 2,3$, then the atlas is again anomaly free.

\begin{table}[ht!]
    \centering
    \begin{tabular}{|c  c |c c c |} \hline
     \textbf{BM} &\textbf{Model} &  $\mathbf{Q'_e}$  & $\mathbf{Q'_{\mu}}$ &  $\mathbf{Q'_{\tau}}$ \\ \hline
      BM1 & $L^{1}_{0,1}$   & 0 & 1 & -1 \\  
      BM2 & $L_{-1,2}^6$   & $-1/6$ & $1/2$ & $-1/3$ \\  
      BM3 & $L_{-2,-1}^4$   & $-1/2$ & $1/4$ & $1/4$ \\  
      BM4 & $L_{1,19}^{20}$   & $-1/20$ & $1$ & $-19/20$ \\  \hline 
    \end{tabular}
    \caption{Benchmark models and their charge assignments, being $L^{n}_{\rho \nu}$ with $n$ the common factor for the charge parameterisation of Eq.~\eqref{Atlas parameterisation}.}
    \label{tab: Benchmark models}
\end{table}

Let us consider four benchmark models (shown in Table. \ref{tab: Benchmark models}), identified not only by a unique choice of $\rho, \vartheta$, but also by a common denominator by which a model can differ from other models with the same parameters~\footnote{This forces the charges to have fractional values (in order to respect perturbativity of the couplings, which are taken to be $\lesssim 1$). Notice that if we set only the values of $\rho$ and $\vartheta$ and require a fractional value for the charges, the model is not specified, since $\rho = 2 \vartheta$ might be a model with $(\rho, \vartheta) = (1/2, 1/4), (1/3, 1/6), $ and so on. We specify the model by this new label $n$}:

\begin{equation*}
 L^{n}_{\rho,\vartheta} = \left \{ Q'_{e}, Q'_{\mu}, Q'_{\tau} \right\}  = \left \{ \frac{\rho}{n}, \frac{\nu - \rho}{n}, -\frac{\nu}{n}\right\},
\end{equation*}

where the superscript $n$ denote the common denominator. The charges were chosen to contrast different scenarios. The Benchmark Model 1 (BM1) corresponds to canonical $L_{\mu} - L_{\tau}$ leptophilic model, largely discussed in the literature~\cite{Heeck:2022znj,heeck2011viable,bauer2021confirming}, and for this reason taken as a main benchmark point.  The second model (BM2) is reminiscent to the electric charges in the quark sector, otherwise is just a representative of a simple choice of the parameters where each family has a different charge. The third Benchmark Model (BM3) is chosen with a larger (twice) value for the charge of the electron with respect to that of the muon. One might expect this  model to be easily excluded since the experimental constraints (e.g. $(g-2)_{e}$, neutrino electron scattering or neutrino trident production~\cite{Altmannshofer:2019zhy, Lindner:2018kjo}) are stronger for electrons. Notice also that this model presents an equal value of the charge of muon and tau-lepton. Finally, Benchmark Model (BM4) is a limiting case of three couplings, with one of them considerably suppressed, i.e. $L_{e} = -\delta, L_{\mu} = 1, L_{\tau} = - (1 - \delta)$, with its similarity to BM1 ($L_{e} = 0, L_{\mu} = 1, L_{\tau} =-  1 $), we can observe similar phenomenology to $L_{\mu} - L_{\tau}$ without needing to justify the complete absence of the coupling with the electron when $\delta \ll 1$.

\subsection{Dark sector in the atlas}

Now we proceed to the introduction of a dark sector in the model. As we discussed before,  a dark matter fermion, with vector-like couplings to the mediator, would not appear in the anomaly cancellation conditions. Therefore it will not affect the charges of the leptons, and thus its coupling to the $Z'$ boson would be independent of $\rho$ and $\vartheta$\footnote{However, if it were axial, with ${Q'}_{\chi_L}, {Q'}_{\chi_R}$, then Eqs.~\eqref{Chiral anomalies d} and~\eqref{Gravitational anomaly} would include these terms, and additional dependencies on the value of the charges would be required. We will not discuss such scenario in the present work, since the dark fermion coupling would be assumed vector-like, with an arbitrary charge.}. 

The interacting Lagrangian of the model, with both SM leptons $\psi$ and dark matter (DM) $\chi$ coupled to the $Z'$ is the following

\begin{equation}
    \mathcal{L}  \supset \sum_{\text{leptons }} Q'_i g' Z'_{\mu} \overline{\psi}_i \gamma^{\mu} \psi_i +  Q'_{\chi} g' Z'_{\mu} \overline{\chi} \gamma^{\mu} \chi,
    \label{Fermion lagrangian}
\end{equation}
being $Q'_{\chi}$ the charge of the dark matter fermions, independent of $\rho$ and $\vartheta$, as mentioned. The choice of fermions as dark matter particles relies on the fact that they offer more mechanisms for stabilization, such as through the behavior of chiral anomaly cancellation in a gauged $U(1)$ leptophilic symmetry.

The first term in the interaction Lagrangian~\eqref{Fermion lagrangian} corresponds to a charged vector-like interaction between the charged leptons and left- and right-handed neutrinos. We will see in Sec.~\ref{section: Muon g-2} how this term is responsible for the muon and electron $(g-2)_{e}$ and how it is constrained by accelerators and cosmological considerations. We will see also that these considerations constrain the mass of the $Z'$ boson to be $\mathcal{O}(10~\mathrm{MeV})$~\cite{Heeck:2022znj}, which is consistent with constrains from neutrino cosmology~\cite{Akita:2023iwq, Abazajian:2019oqj}, and accelerator searches of $Z'$ bosons~\cite{cms2021search,atlas2021search,cdf2012search}.

The second term of Lagrangian~\eqref{Fermion lagrangian} is the vector-like coupling of the dark matter candidate $\chi$ to the same gauge boson $Z'$, with charge $Q'_{\chi}$. It has three phenomenological consequences. The first one is the production of high energy cosmic rays from DM annihilation to electrons, which opens the possibility of an indirect detection channel and whose detailed treatment is beyond the scope of this work. Current searches, however, set bounds on massive vector bosons above $\mathrm{GeV}$ order to baryonic final states~\cite{Chang:2018bpt, HAWC:2018szf, Bartlett:2022ztj, VERITAS:2017tif}, and above $1~\mathrm{TeV}$ to final lepton states~\cite{VERITAS:2017tif, Elor:2015bho, Guo:2023axz, IceCube:2023ies}.  Similarly, annihilation of dark particles into the baryonic sector during the early universe determine the current (relic) abundance of dark matter, as we will discuss in section~\ref{section: Production of Dark Matter Particles}. Finally, since the model contains a mediator $Z'$ and a dark matter candidate $\chi$ that interacts with itself via this mediator, this framework encodes a plausible SIDM sector. The details of such framework will be discussed in section~\ref{section: Small Scale Effects}, in the context of the mismatch between CDM simulations and observations of small scale haloes~\cite{2017ARA&A..55..343B, Kaplinghat:2015aga, Tulin:2013teo}.

The dark fermion $\chi$ and the $Z'$ boson itself have standard mass terms assumed to be generated via a Stückelberg mechanism~\cite{Feldman2007Phenomenology}, where a heaviest Higgs boson decouples from the scalar spectrum responsible of the spontaneous symmetry breaking into the $U'(1)$ symmetry. Hence our model can be seen as an effective approach taking only into account the interactions with effects on the phenomenology of dark matter sector and the non-universal behavior for the flavor in the lepton sector.  Other details of the mass generation of the complete model and of the underlying scalar sector will not be discussed here, neither will be the neutrino mass generation mechanism of the model. For similar works that do discuss neutrino mass generation and more sophisticated scalar sectors with canonical spontaneous symmetry breaking, see~\cite{Heeck:2022znj, Lou:2024fvw, DiBari:2024jkj, Abdallah:2024npf, Bittar:2024ryj}.

Before we proceed let us make a comment on the possible $Z'-Z$ mixing that arises in our model. Terms $\propto c_{12} Z'^{\mu \nu} Z_{\mu \nu}$ do not vanish in leptophilic extensions because they are gauge invariant. Besides, even if this term is  zero in the lagrangian, loop effects can produce a net non-zero value \cite{Babu:1997st, Drees:2021rsg}. For $L_{\mu} - L_{\tau}$ models, the kinetic mixing at vanishing momentum transfer $c_{12} \approx g'/70$. We therefore neglect this contribution in the diagrams for the forthcoming discussion. For details regarding the mixing mechanism and its phenomenological consequences on $L_{\mu} - L_{\tau}$ models, see~\cite{Ekstedt:2016wyi, LoChiatto:2024guj, Ferdiyan:2024jko}.

In the following sections, we describe the constraints on our models from a selected set of phenomenological facts where leptophilic couplings that emerged from the atlas have a strong impact.

\section{Muon g-2}\label{section: Muon g-2}

The muon $g-2$ experiment at FermiLab National Laboratory (FNAL) reported a new measurement of the anomalous magnetic moment of the muon (denoted as $(g-2)_{\mu}$) with increased precision and statistics with respect to previous reports~\cite{Muong-2:2006rrc, Muong-2:2021ojo}. The current global average is set $5.1\sigma$ from the SM prediction~\cite{Muong-2:2023cdq}:

\begin{equation}
    \Delta a_{\mu} = 249(48)\times 10^{-11}.
    \label{Delta a_mu}
\end{equation}

Even when the latest measurement set a discrepancy beyond the $5\sigma$ benchmark for claiming discovery of new physics, it is not yet the case for the $g-2$ muon discrepancy~\cite{Aoyama:2020ynm}. This is because there is an existent tension in the determination of the Hadronic Vacuum Polarization correction to the vertex~\cite{Borsanyi:2020mff}. The two methods employed in the $g-2$ Theory Collaboration (data-driven and lattice-QCD) are $3.9\sigma$~\cite{Ce:2022kxy} away from one another. This discrepancy makes the reported value insufficient to claim a new physics discovery since the theoretical value computed with Lattice-QCD is closer to the experimental value. Further analysis in the hadronic part of the theory will determine ultimately if this experiment is finally either revealing BSM physics or not.

\begin{figure}
    \centering
    \includegraphics[width=0.8 \textwidth]{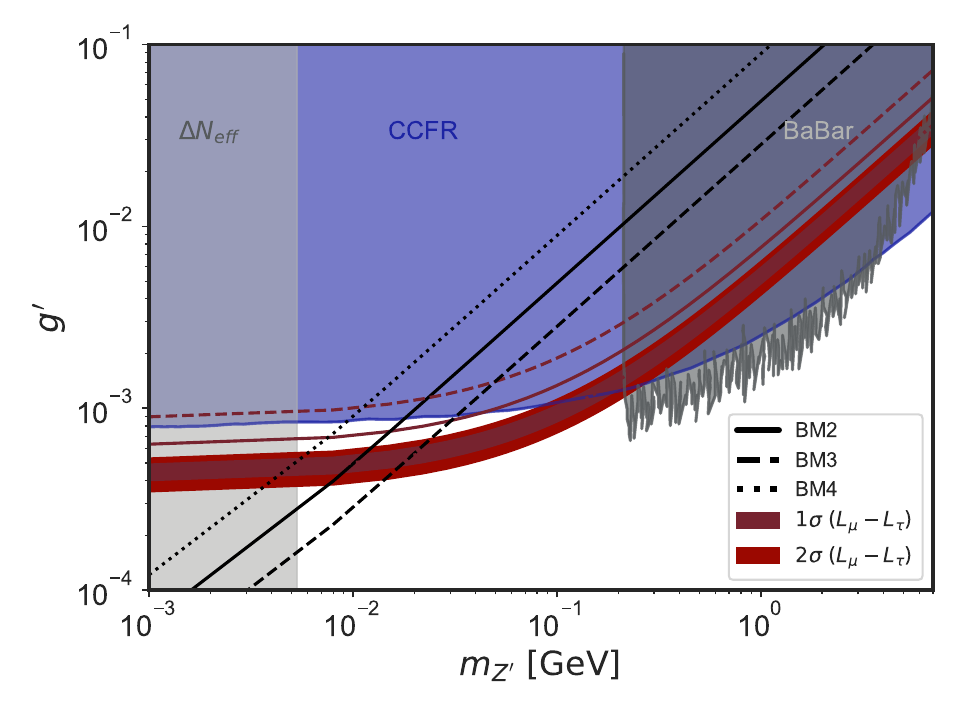}
    \caption{Parameter region favoured by the $(g-2)_{\mu}$ discrepancy at $1$ and $2\sigma$ for BM1 (red bands) and only central value for the other models of the atlas in Tab.~\ref{tab: Benchmark models} (red lines). Also it is depicted the prediction for $(g-2)_e$ for the same models (black lines), and the experimental constraints from $\Delta N_{eff}$~\cite{Kamada:2015era}, \textsc{CCFR} measurements in neutrino tridents~\cite{CCFR:1991lpl} and BaBar searches of $Z'$ mediators~\cite{BaBar:2016sci}. Solid, dashed and dotted lines correspond to BM2, BM3 and BM4 respectively for both muons an leptons.}
    \label{fig: g-2 in the atlas}
\end{figure}

In the meantime, theoretical efforts continue and consider the current status as baseline for the feasible presence of new particles explaining the anomaly. One of the most studied of such new physics scenarios is the exchange of a $Z'$ boson in the one loop vertex of the muon and photon, as a next to leading correction to the anomalous magnetic moment~\cite{buen2022status,altmannshofer2022explaining}. In the case of our model with the leptophilic atlas, such coupling exists but so does the one with the other leptons ($e$ and $\tau$), in each case with a contribution given by

\begin{equation}
     a_{l}^{Z'} = \frac{{g'}_l^2 }{8\pi^2}\int_0^1 \frac{2m_l^2 x^2(1-x)}{x^2 m_l^2 + (1-x)m_{Z'}^2} dx,
     \label{Z' contribution to g-2}
\end{equation}
where $g'_l = g' {Q'_l}$ and $l = e, \mu, \tau$. The mass of the mediator is identified as $m_{Z'}$. In the case of the electron, the anomalous magnetic moment has been measured with higher precision and agrees with the SM calculation by $2.4\sigma$~\cite{Hanneke:2008tm}:

\begin{equation}
    \Delta a_e = -87(36) \times 10^{-14};
    \label{Delta a_e}
\end{equation}
hence, we can use the discrepancy as a constraint for the parameter space of the model. In Fig.~\ref{fig: g-2 in the atlas}  we show the preferred region in the $(g'_{\mu}, m_{Z'})$ plane for all benchmark scenarios presented in Tab.~\ref{tab: Benchmark models}. The red bands are the $1$ and $2\sigma$ for BM1, while for the other models we show only the central value, as indicated (in red as well). The black diagonal lines refer to the corresponding central value of the anomalous magnetic moment of the electron, which, as commented, we can interpret (from that line upwards) as a constraining region for the value of $g'$ at given $m_{Z'}$ (there is no associated line to BM1 since in this model $Q'_e = 0$, see Tab.~\ref{tab: Benchmark models}). We also show in Fig.~\ref{fig: g-2 in the atlas} other constraints for the muon-$Z'$ coupling that would contribute to in $e^+ e^- \to \mu^+ \mu^- \mu^+ \mu^-$ from BaBar searches~\cite{BaBar:2016sci}, CCFR studies on lepton trident production~\cite{CCFR:1991lpl}, and $\Delta N_{\text{eff}}$ cosmological constrains on light mediators~\cite{Kamada:2015era}. 

Notice also from Fig.~\ref{fig: g-2 in the atlas}, that BM3 is, as expected, severely constrained due to the overlapping with CCFR measurements, $N_{eff}$ and BaBar exclusion regions. The framework BM2 has a very small window and we see that the $(g-2)_{e}$ constraint plays an important role in almost ruling it out. On the other hand, BM4 differs from BM1 only in a region of $m_{Z'} \sim 5 - 8 ~\mathrm{MeV} $, which was expected to be so given the couplings presented in Tab. \ref{tab: Benchmark models}.

\section{Production of Dark Matter Particles}\label{section: Production of Dark Matter Particles}

We turn now our attention to discuss the dark matter sector of the model. For the dark candidate, we assume a Dirac fermion $\chi$ charged with $Q'_{\chi}$; which we may take equal to one without lost of generality, since this sector is not related to the leptophilic atlas. 
The $Z'$ boson is a portal between the visible and the dark sectors because it is coupled simultaneously to both, as is manifest in the Lagrangian of Eq.~\eqref{Fermion lagrangian}.   In the early universe, such portal allows a thermal equilibrium between SM particles and the DM candidate provided that the temperature of the thermal bath $T$ satisfies $ m_{\chi} \lesssim T$ ($m_{\chi}$ is the dark fermion mass). When the universe expands and cools down, the process $\chi \overline{\chi} \to l^+ l^-$ ($l$ encodes all the SM leptons) overcomes its inverse $l^+ l^- \to \chi \overline{\chi}$, and the density of dark matter diminishes exponentially. This effect is however only one annihilation channel, the other being $\chi \overline{\chi} \to Z' Z'$.  The thermal cross sections for these processes are given by~\cite{Altmannshofer:2016jzy}

\begin{align}
    \langle \sigma v \rangle (\chi \overline{\chi} \to l^+ l^-) &= \frac{{g'}_l^2 {g'}_{\chi}^2}{2\pi} \sqrt{1 - \frac{m_l^2}{m^2_{\chi}}} \frac{2m_{\chi}^2 + m_l^2}{(4m_{\chi}^2 - m_{Z'}^2)^2}, \label{Thermal cross section to leptons} \\
    \langle \sigma v \rangle (\chi \overline{\chi} \to Z' Z') &= \frac{{g'}_{\chi}^4}{16 \pi m_{\chi}^2} \left( 1 - \frac{m_{Z'}^2}{m_{\chi}^2}\right)^{3/2}\hspace{-0.2cm}\left( 1 - \frac{m_{Z'}^2}{2 m_{\chi}^2}\right)^{-2}\hspace{-0.4cm}. \label{Thermal cross section to Z's}
\end{align}
$g'_{\chi}$ is the coupling constant between the dark fermions $\chi$ and the portal $Z'$. Both processes contribute to the reduction in dark matter density during the early universe, but the cross section of Eq.~\eqref{Thermal cross section to Z's} is the dominant annihilation channel, i.e., when $m_{Z'} \leq m_{\chi}$. This hierarchy is a condition for the stability of the gauge boson; therefore, we will consider only this channel of annihilation. 
As the universe continues its expansion, dark matter particles eventually stop annihilating, their density remains roughly constant and we say that the dark sector \textit{freezes-out}. 
This effect is mathematically described by the solution of the Boltzmann equation, which is given as follows

\begin{equation}
    \frac{dY}{dx} = - \left( \frac{45}{M_p^2\pi}\right)^{-1/2} \frac{g_{*}^{1/2} m_{\chi}}{x^2} \langle \sigma v \rangle (Y^2 - Y^2_{eq}) ,
    \label{Boltzmann equation}
\end{equation}
where $M_{p}=1.22 \times 10^{19}$ GeV is the Planck Mass and $g_{*}$ is the relativistic number of degrees of freedom for energy density~\cite{Kolb:1990vq} The variable $x = m_{\chi}/T$ and $Y = n/s_{\gamma}$ is the \textit{yield}, a comoving measure of the number density, whose value at equilibrium is given by
\begin{equation}
    Y_{eq} = \frac{n^{\text{eq}}}{s} = \frac{45}{2\pi^4} \frac{g_{*}}{g_{*S}}\left(\frac{m_{\chi}}{T}\right)^{3/2} e^{-m_{\chi}/T}.
\end{equation}  
$g_{*S}$ refers to the effective number of entropy degrees of freedom. One can solve Boltzmann equation numerically with the thermal cross section given by Eq.~\eqref{Thermal cross section to Z's} and use the comoving number density after freeze-out to estimate the density of dark matter particles today, given by~\cite{Lin:2019uvt}:

\begin{equation}
    \Omega_{\chi 0} h^2 = \frac{\rho_{\chi}^0h^2}{\rho_{c,0}} = \frac{m_{\chi} s_{\gamma} Y_0 h^2}{\rho_{0,c}},
    \label{Relic density today}
\end{equation}
where $\Omega_{\chi 0}$  means the current value of the relic density, $\rho_c$ is the critical density and $h$ is the Hubble constant in units of $100~\mathrm{km/s/Mpc}$. The relic density has been measured by PLANCK satellite~\cite{2020A&A...641A...6P}:

\begin{equation}
    \Omega_{\text{DM}} h^2 = 0.1199\pm 0.0027,
\end{equation}
at $68\%$~$\mathrm{C.L}$. We can use this cosmological observable, assuming that the dark fermion $\chi$ (and its anti-particle) constitute all the dark matter in the universe, to determine the required coupling constant for a given DM mass (or viceversa), independently of the mediator mass. Although, the independence on $m_{Z'}$ is not exact. In our model, the independence on mediator mass comes from the ratio $m_{Z'} / m_{\chi} \sim 10~\mathrm{MeV} / 100~\mathrm{GeV} \sim \mathcal{O}(10^{-4})$, so we can neglect it in Eq.~\eqref{Thermal cross section to Z's}. In the following section we will discuss explicitly  the common regime in which this mechanism populates all the dark matter in the universe and the values the parameters of our model need to take to successfully accomplish this scenario (see Fig.~\ref{fig: Parameter Space Plot}).

Since we have discussed the production mechanism of dark matter particle $\chi$ through thermalization, we will consider in the following section how the self-interacting properties mediated by the $Z'$ boson can alleviate the so-called core-cusp problem, as we will discuss in the following section. Moreover, we will find a region in the parameter space for ${g'}_{\chi}$ and $ m_{\chi}$ compatible with the results of the present section, capable of accounting for the cross sections required to form cores inside of small galactic halos.

\section{Small Scale Effects}\label{section: Small Scale Effects}

It has been already studied in the literature the possible contribution of $U(1)_{L_{\mu} - L_{\tau}}$ models with a dark sector to alleviate the core-cusp problem by the introduction of a  fermion or scalar field (a SIDM candidate) that couples to the mediator gauge boson~\cite{Heeck:2022znj, Kamada:2018zxi}. Hence, one crucial aim for our approach is to study the behavior of dark sector with a mediator emerging from the extended leptophilic model confronting the challenges of CDM in small scales. 

At the fundamental level, the interaction between dark matter particles acts in such a way that cross section depends on their relative velocity, which is intimately related to the length scale of the astrophysical object. The cross sections per unit of dark matter mass are smaller for clusters scale of order $\sigma/m_{\chi}<0.1$ cm$^2/$g, with velocities of $v>10^{3}$ km$/$s~\cite{Andrade:2020lqq}. Instead, the cross sections per unit of mass for dwarf galaxies are in the range $1$ cm$^2/$g $<\sigma/m_{\chi}<100$ cm$^2/$g  for velocities $v<10^2$ km/s~\cite{Correa:2020qam}. 

The mediator in the dark sector regulates the amount of transfer momentum in the interaction, that is the origin of the velocity dependence of the process. Furthermore, the exchange of a light mediator is the feasible feature to explain: 1. Core formation of dwarf galaxies~\cite{Elbert:2014bma}, 2. Less dense halos of ultra-diffuse galaxies~\cite{Nadler:2023nrd}, 3. The observed anti-correlation between central dark matter densities and the orbital pericenter distance of Milky Way's dwarf spheroidals~\cite{Correa:2020qam}, and 4. Other so-called small scale problems, e.g., missing satellites and too big to fail\footnote{The missing satellites problem is originated from an overabundance of DM sub-halos in only dark matter simulations with respect to what has been observed around the Milky Way. The too big to fail problem~\cite{2017ARA&A..55..343B}  relies in  the lack of observations of galaxies inside sub-halos with $M_{\text{vir}} = 10^{10} ~\mathrm{M}_{\odot} $ around the Milky Way, which according to simulations should be a feature in a CDM scenario~\cite{Tulin:2017ara}.}~\cite{2017ARA&A..55..343B,Tulin:2017ara}.

At the same time the model resembles the cold dark matter behavior at large scales, given the velocity dependence of the self-interaction cross section confronting the constraints of galaxies~\cite{Elbert:2014bma}, from which the cross section also decreases with velocity, obtaining the same behavior of the expected for CDM models at those scales. In what follows, we want to explore quantitatively whether our model can produce the cross sections required by simulations to account density profiles in galaxies (proxies of the circular velocities) that defy the standard collisionless cold dark matter (CDM).

\begin{figure}
    \centering           
    \includegraphics[width = 0.8\textwidth]{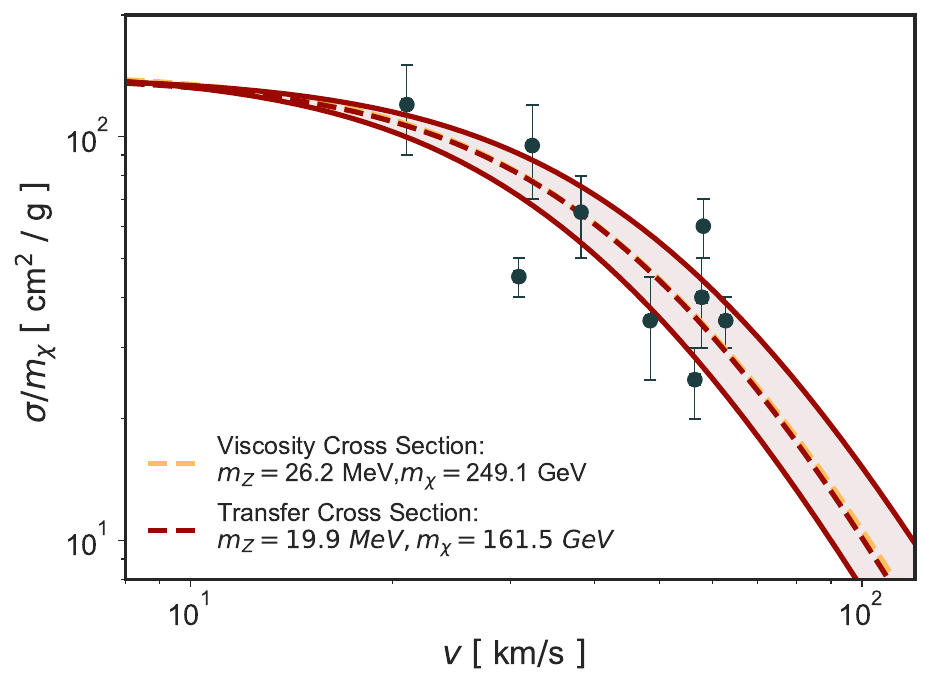}
    \caption{Fit result for the Milky Way satellites dwarf spheroidals (from left to right: Leo II, UM, Sextants, Cvn I, Carina, Draco,Fornax,  Leo I and  Sculptor ) using the viscosity cross section of Eq.~\eqref{eq:tsigmav}) and transfer cross section of Eq.~\eqref{eq:tsigmat}. The data is originally performed in~\cite{Correa:2020qam}. In the fitting procedure we do not consider the UM DSph (the outlier), because this data point yields a strong tension in the optimal parameters combination regarding the general tendency of the original target function.}
    \label{fig:dSphfit}
\end{figure}

To begin, we take the astrophysical data in halos of Milky Way dwarf satellites~\cite{Correa:2020qam} to test the model regarding the behavior of cross section per unit mass as a function of the velocity of DM particle.  Likewise, we use the data for a thermal average cross section in astrophysical objects Low-Surface-Brightness and Clusters of galaxies to confront the model prediction to the self-interacting velocity-weighted cross section\footnote{The fitted function is a halo model that consists on a NFW part at large radii and an isothermal profile formed in the core of the galactic structure. The core formation in the galactic structure can be understood in a phenomenological way through the isothermal profile that emulates properly the density behavior obtained by galaxy simulations involving self interactions~\cite{Elbert:2014bma}. Instead, the density for the outer part of the halo follows a NFW density profile, encoding the initial conditions and thermalization mechanisms for the halo~\cite{Zentner:2022xux}.}~\cite{Kaplinghat:2015aga}. Furthermore, we use the data obtained from the halo-formation model~\cite{Correa:2020qam}, that reconstructs a velocity averaged cross section for simulated halos that match central density and velocity of some of the Milky Way satellite galaxies. 

The SIDM that we have been discussing acts through a mediator (which in the general case can be vector or scalar), in the form of a Yukawa potential in a non relativistic limit, i.e., $V(r)=\pm\alpha_{\chi}e^{-m_{\phi}r}/r$. When the model parameters $\alpha_{\chi}, m_{\chi}, m_{Z'}$ (being $\alpha_{\chi} = {g'}_{\chi}^2/4\pi$) fit in the Born regime, given by $\alpha_{\chi} m_{\chi} \ll m_{Z'}$, we have for the differential cross section~\cite{Tulin:2013teo}:

\begin{equation}
\frac{d\sigma}{d\Omega}=\frac{\sigma_{0}}{4\pi} \frac{1}{\left[1+\frac{v^2}{w^{2}}\sin^{2}\left(\frac{\theta}{2}\right)\right]^{2}},
\end{equation}
where $\sigma_{0}=4\pi \alpha_{\chi}^{2}m_{\chi}^{2}/m_{Z'}^{2}$ is the geometric cross section and $w=m_{Z'}c/m_{\chi}$ is the mass fraction times the speed of light $c$.

The target models to perform the fits are based on the viscosity ($\sigma_{V}$) and transfer ($\sigma_{T}$) cross sections. The viscosity cross section arises from kinetic theory with dissipative effects in the interactions of fluids. It is introduced because it yields a better figure of merit to interpret the results of simulations based on gravothermal catastrophe processes as those presented in~\cite{Correa:2020qam}, since they are driven out with heat transfer effects. The functional form of $\sigma_{V}$ is determined by

\begin{equation}
    \sigma_V = \int \sin^{2}{\theta} \frac{d\sigma}{d\Omega} d\Omega.
    \label{Viscosity cross section}
\end{equation}

On the other hand, the transfer cross section $\sigma_{T}$ shows the behavior regarding the distribution of momenta in the phase space, i.e., interactions that lead to a large amount of momentum transfer have a higher weight and interactions with smaller momenta are suppressed. The $\sigma_{T}$ is similarly determined by

\begin{equation}
\sigma_{T}=\int (1-\cos\theta) \frac{d\sigma}{d\Omega} d\Omega.
    \label{Transfer Cross section}
\end{equation}

The target cross sections to the fits of the Milky Way Satellite (MWS) galaxies are the following functions~\cite{2022JCAP...09..077Y} 

\begin{subequations}
\begin{align}
\sigma_{V}&= \displaystyle \frac{6{\sigma }_{0}{w}^{6}}{{v}^{6}}\left[\left(2+\displaystyle \frac{{v}^{2}}{{w}^{2}}\right)\mathrm{ln}\left(1+\displaystyle \frac{{v}^{2}}{{w}^{2}}\right)-\displaystyle \frac{2{v}^{2}}{{w}^{2}}\right], \label{eq:tsigmav}\\
\sigma_{T}&=\frac{2\sigma_{0}\omega^{4}}{v^{4}}\left[\ln\left(1+\frac{v^{2}}{w^{4}}\right)-\frac{v^{2}}{v^{2}+w^{2}}\right]\label{eq:tsigmat},
\end{align}
\label{target cross section functions}
\end{subequations}
divided by the mass of $\chi$. The scaling of the cross section $\sigma_{0}$ and the mass fraction $\omega$ are described by means of
\begin{align*}
\sigma_{0}& =275.7 \left(\frac{\alpha_{X}}{0.01}\right)^{2}\left(\frac{m_{\chi}}{10 \text{ GeV}}\right)\left(\frac{10 \text{ MeV}}{m_{Z'}}\right)^{4},\\
w &= 300 \left(\frac{m_{Z'}}{10 \text{ MeV}}\right) \left(\frac{10 \text{ GeV}}{m_{\chi}}\right) \hspace{0.1cm}\frac{\text{km}}{s}.
\end{align*}

\begin{figure}
    \centering
    \begin{subfigure}[b]{0.30\textwidth}
    \includegraphics[width=\textwidth]{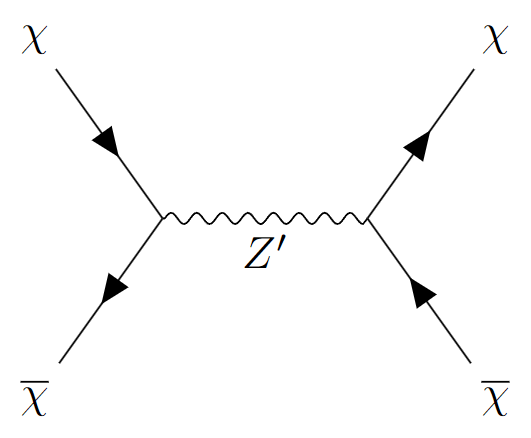}
    \end{subfigure}
    \begin{subfigure}[b]{0.30\textwidth}
    \includegraphics[width=\textwidth]{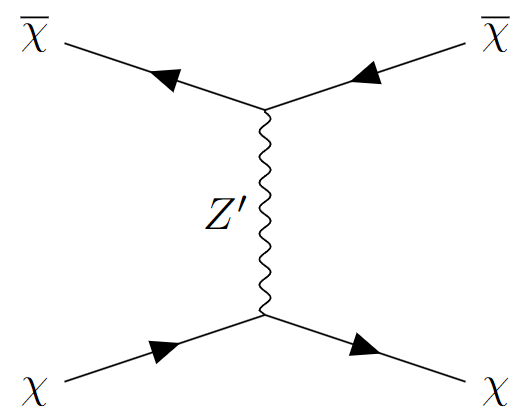}
    \end{subfigure}
    \caption{Feynman diagrams for $s-$ and $t-$ channels in $\chi \Bar{\chi} \rightarrow \chi \Bar{\chi}  $ scattering}
    \label{fig:DM diagram for chi chibar -> chi chibar}
\end{figure}

To find the optimal parameters combination in the search for the best fit confronting the MW dwarf satellites $\sigma/m_{\chi}$ as a function of the dispersion velocity, we run a Markov Chain Montecarlo (MCMC) with $N_{\text{walkers}}=48$, $N_{\text{steps}}=10000$ (with $2000$ steps of optimization). Figure~\ref{fig:dSphfit} shows the data from Ref.~\cite{Correa:2020qam} and our best fitted function with best fit parameters (fixing the coupling $\alpha_{X}$ to be $0.01$), in the case of the viscosity cross section (with a reduced $\chi^2$ of $2.49$):

\begin{equation}
    \begin{aligned}
        m_{Z'} &= 25.1^{+1.1}_{-1.2} \text{ MeV},\\
        m_{\chi} &= 209.6 ^{+39.5}_{-39.4}\text{ GeV}.
    \end{aligned}
    \label{Best fit parameters vicosity}
\end{equation}

In the case of transfer cross section, the best fitted parameters are the following (with a reduced $\chi^2$ of $2.48$):

\begin{equation}
    \begin{aligned}
        m_{Z'} &= 19.9^{+0.9}_{-0.9} \text{ MeV},\\
        m_{\chi} &= 161.5 ^{+25.5}_{-28.3}\text{ GeV}.
    \end{aligned}
    \label{Best fit parameters transfer}
\end{equation}

Now, we proceed to evaluate the behavior of our parameters in front of a velocity averaged cross section, that is a typical indicator of SIDM effects obtained from data of the different astrophysical objects of galactic and clusters scales. To achieve this comparison, equations~\eqref{target cross section functions} are then numerically integrated from $v_0 = 0$ to $v_0 = v_{max} $ (which we take to be two standard deviations of the Maxwell-Boltzman distribution: $2\sqrt{v_0^2 (3\pi - 8)/\pi}$) over the relative velocity to get a thermal average cross section:

\begin{equation}
    \langle \sigma_{T(V)} v \rangle = \frac{1}{(2\pi v_0^2 )^{3/2} } \int \sigma_{T(V)} v e^{-\frac{1}{2}v^2/v_0^2}d^3 v .
    \label{Velocity averaged cross section}
\end{equation}

The velocity averaged cross section is fitted by using a larger data set consisting of both sets of data described above that contain information of clusters, isolated dwarf irregulars and low surface brightness galaxies. 

\begin{figure}
    \centering
    \includegraphics[width=0.8\textwidth]{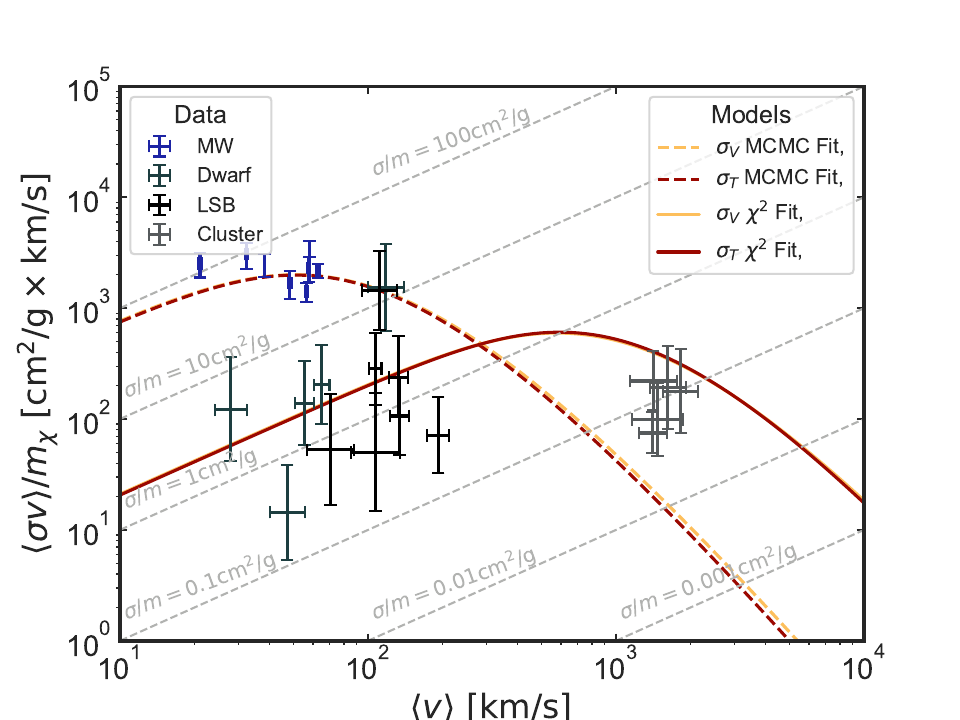}
    \caption{Thermal cross section per unit of mass as a function of the average particle velocity in a halo. Data points are derived from different astrophysical objects such as MW satellites~\cite{Correa:2020qam}, Dwarfs from Little Things, Low Surface Brightness galaxies, and groups of clusters~\cite{Kaplinghat:2015aga}. The solid lines represent the functions evaluated in best fit parameters combination for the viscosity (red) and transfer cross section (yellow). The gray dashed lines represent constant values for cross section over the DM mass, corresponding to the expected values for the different scales of the astrophysical objects. }
    \label{fig:GalaxyFit1}
\end{figure}

For the case of the numerical function in Eq.~\eqref{Velocity averaged cross section}, we performed a standard $\chi^2$-fitting procedure\footnote{The $\chi^2$ fit proves to be more computationally efficient in this case than the MCMC fit, due to the complexity of the integral.} to the data from Refs.~\cite{Correa:2020qam} and~\cite{Kaplinghat:2015aga}. The best fit parameters are used to plot the transfer and viscosity cross sections in Fig.~\ref{fig:GalaxyFit1}  (solid lines in red and yellow, respectively). The best fit parameters, with coupling set to $\alpha_{\chi} = 0.01$, are $  m_{Z'} = 22.72 ~ \mathrm{ MeV}$ and $m_{\chi} = 24.08~\mathrm{ GeV}$, for the viscosity cross section. For the transfer cross section, the best fit parameters are $m_{Z'} = 22.47  ~ \mathrm{ MeV}$, and $m_{\chi} = 22.53 ~\mathrm{ GeV}$. We also show the same viscosity (transfer) averaged cross section with the parameters given in Eq.~\eqref{Best fit parameters vicosity} (accordingly in Eq.~\eqref{Best fit parameters transfer} ) and found through the MCMC procedure in the yellow (red) dashed lines. 

The data in the upper left side of Fig.~\ref{fig:GalaxyFit1} (blue dots with vertical error bars) correspond to the same set of (Milky Way Satellites) MWS galaxies from the data used in Fig. \ref{fig:dSphfit}, now in terms of a thermal cross section. Besides crossing the MWS data points, unlike in the $\chi^2$-best fit parameter case represented by the solid lines, the dashed lines also show a steeper descent in their large scale behavior, with respect to the solid lines. The decreasing tendency of the function evaluated in best-parameters fit appears to disagree with the expectation of cluster data and simulations, as seen in the Fig.~\ref{fig:GalaxyFit1}. However, the large scale limit establishes only a restriction in how high a cross section can be, instead of determining a fixed value; bringing the model closer to a CDM scenario (which is desirable in this regime), when the cross sections are smaller, like for the model evaluated in the best fit parameters of Eqs.~\eqref{Best fit parameters vicosity} and~\eqref{Best fit parameters transfer}. Furthermore, we would like to point out the fact that the mass of the mediator for both fitting procedures are rather similar (around $m_{Z'} \sim 20~\mathrm{MeV} $, consistent also with the results of the previous sections), whereas the prediction for the mass of the DM particles varies in one order of magnitude. The fact that both fits lead to different parameter values comes from the different physical processes involving the two astrophysical objects and determining, for example, their dynamics, formation and evolution (not to mention the huge differences in mass and size)\footnote{Dwarf galaxies are considered to be among the first structures to form from the density inhomogeneities in the CMB~\cite{Dayal:2018hft}, while clusters of galaxies form later in the cosmic timeline through the hierarchical merging of structures. Dwarf halo dynamics might be influenced by tidal stripping caused by nearby (larger) galaxies, supernova feedback in their interior, and reionization of the interstellar medium~\cite{Muni:2024jts}. On the other hand, clusters of galaxies are influenced by the gravitational interactions between the member galaxies~\cite{Moore:1999gc}.}. On the contrary, the same best fit value for the mediator mass in the two fits marks the spectator nature of that parameter.

\begin{figure}[ht!]
        \centering
        \includegraphics[width=0.8\textwidth]{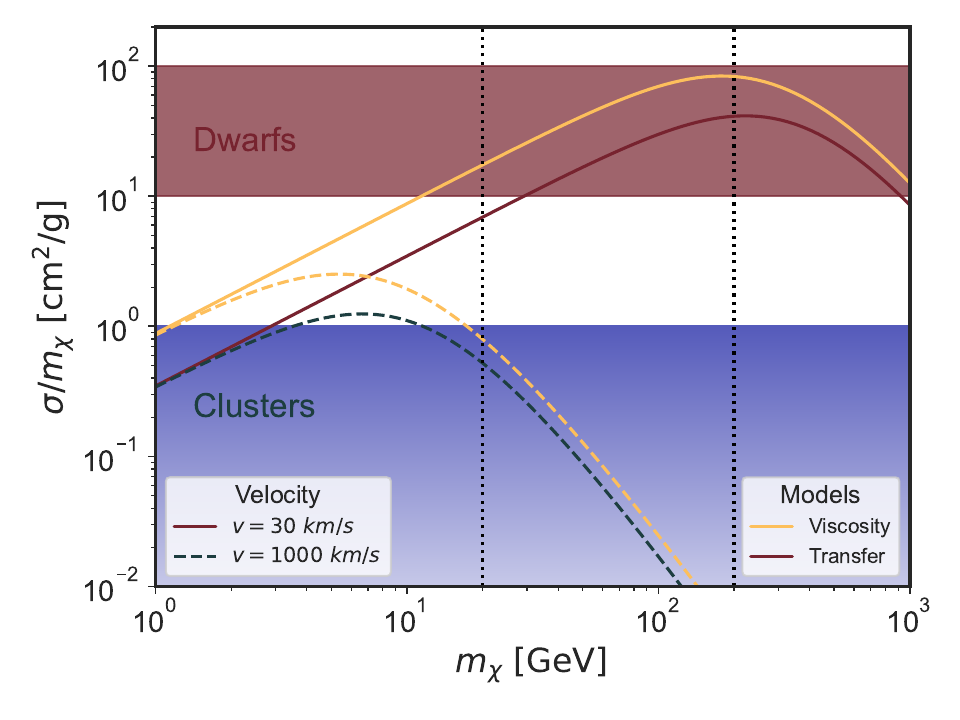}
        \caption{Transfer and viscosity cross-sections per mass as a function of DM particle mass. We show the compatibility of our model with expected cross-section to core-formation in dSph (red band), which also avoids constraints on galaxy clusters (blue-shaded region). The range of compatibility happens when the red solid (blue dashed) curves, corresponding to a velocity $v\approx 10~\mathrm{km/s} $ ($10^{4} ~\mathrm{km/s}$) pass over those regions, for the case of the transfer cross sections. Their counterparts for the viscosity cross section are the yellow lines, with the line-type referring to the different relative velocity.}
        \label{fig: Transfer cross-sections}
    \end{figure}

To see why both values of the mass of the DM fermion given by the two fits are in agreement with large cross sections at low velocity and small cross sections at large velocity, in the amount required by simulations to solve small scale problems, we plotted in Fig.~\ref{fig: Transfer cross-sections} the cross section per mass as function of DM mass for the velocity regimes that would correspond to dwarf galaxies ($v = 30~\mathrm{km/s}$, solid lines) and cluster of galaxies ($v = 1000~\mathrm{km/s}$ for dashed lines) environments, fixing $\alpha_{\chi}$ to the value used for the fits and the $Z'$ mass to $20~\mathrm{MeV}$, as obtained from the fits as well. A wide range for the parameter $m_{\chi}$ is seen to satisfy the condition of $10~\mathrm{cm^2/g} \leq \sigma_i/m_{\chi} \leq 100~\mathrm{cm^2/g} $ (with $i = T, V$), that generates cores inside small scale halos (red shaded  region)~\cite{Correa:2020qam, Elbert:2014bma}, while the constraint from large scale objects dictating $\sigma_i/m_{\chi} \lesssim 0.1~\mathrm{cm^2/g}$ is satisfied (the blue shaded region)~\cite{2021JCAP...01..024S, Sagunski:2020spe}. The dotted vertical lines represent precisely the values given after the fits discussed.

Another interesting feature to observe in Fig.~\ref{fig: Transfer cross-sections} is the proximity between transfer (red and blue lines) and viscosity (yellow lines) cross sections for the two velocity values. We can see a similar behavior for both viscosity and transfer cross section in their mass dependence, except for the order of the cross sections at low dark matter mass; at large DM mass they overlap instead. In Figs.~\ref{fig:dSphfit} and \ref{fig:GalaxyFit1}, we compare as well the viscosity cross section (yellow lines) and the transfer cross section (red lines). The functional similarity of both cross sections holds also in the velocity dependence. We see that they are linked in the light-mediator SIDM model that we consider. Part of the analyses previous to this work consider the transfer cross section as the more approximate figure to hydrodynamic effects of SIDM, also because of its computational effectiveness~\cite{Correa:2020qam}, whereas other rely on more physical considerations and establish that the quantity in charge of describing heat dissipation and distribution over the halo, and therefore determining its density distribution, is the viscosity cross section~\cite{Nadler:2023nrd, Yang:2022hkm}. The analysis presented shows that, even if the second is the case, and the more accurate quantity to rely on dark matter thermodynamics is the viscosity cross section, the effects described in small galactic halos are similar for a wide range of the parameters. 

Finally, we recall that the value for the mediator mass found after the fits, and the one discussed in Sec.~\ref{section: Muon g-2} can be the same: around $20 ~\mathrm{MeV}$ and up to $50~\mathrm{MeV}$. Likewise, the thermal production of the DM fermion yields a relic density value independent of the gauge boson mediating annihilation when $m_{Z'} \ll m_{\chi}$. Such coincidence motivates us to regard the mediator as the same gauge boson of a new $U(1)'$ gauge group under which SM fermions and DM are charged (our leptophilic atlas of Eq.~\eqref{Atlas parameterisation}, for example). Moreover, the non-universality of the $U(1)'_l$ charges of the group allows to have $g'_{\mu} = Q'_{\mu} g'\sim 10^{-4}$, while $g'_{\chi} = Q'_{\chi} g' = 0.35$. The small value in the muon coupling explains the discrepancy in its anomalous magnetic moment determination, respecting constraints on $Z'$ searches, and the  large coupling to $\chi$ is high enough to allow core formation during the evolution of a DM small scale halo, all within the same particle physics model. We highlight that this important feature is new in our model. Previous works in the literature (see~\cite{Heeck:2022znj} and~\cite{Kamada:2015era}) do not consider the same mediator or the same coupling in the discussion of the two problems. In the present case, this is achieved at the expense of a non-universal character of the charges for the different families, which translates in flavor violation in the leptonic sector. 

\begin{figure}[ht!]  
    \centering
    \includegraphics[width=0.8\textwidth]{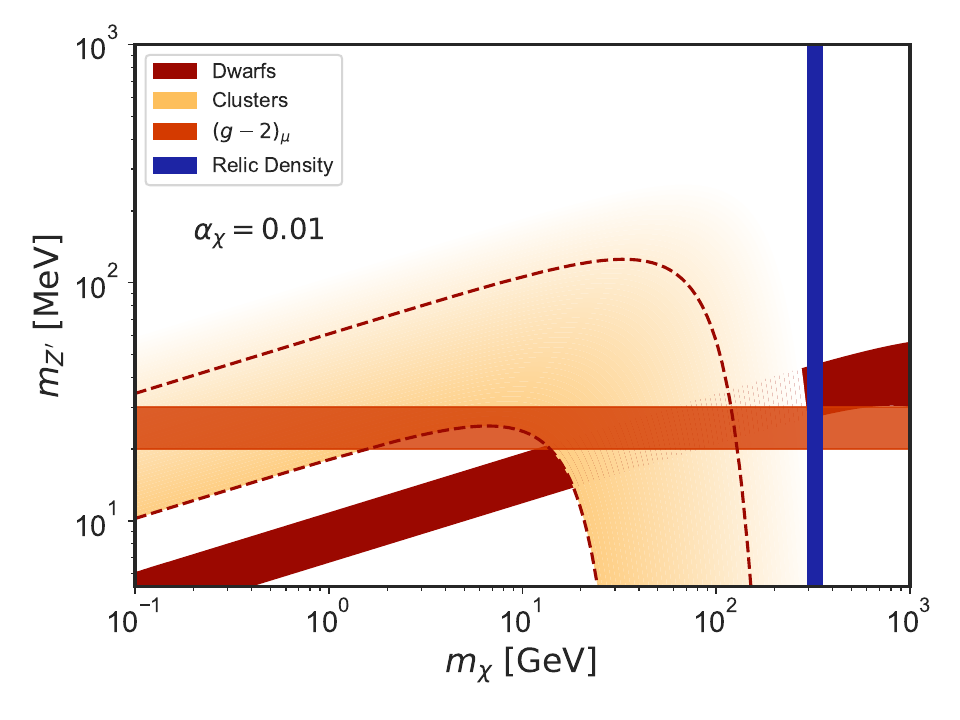}
    \caption{Parameter space of the masses of the DM fermion $m_{\chi}$ and the gauge boson $m_{Z'}$, with a dark coupling of $\alpha_{\chi} = 0.01.$ We show jointly the value preferred to the mass of the mediator from $(g-2)_{\mu}$ for a coupling to the muon ${g'}_{\mu}\sim 10^{-4}$ (see Sec.~\ref{section: Muon g-2}) in orange, the DM mass range to produce the precise amount of DM density (see Sec.~\ref{section: Production of Dark Matter Particles}) in blue, and the values of viscosity (red) and transfer (yellow) cross sections per unit mass $\sigma_{V,T}/m_{\chi}$ in the ranges specified by Fig. \ref{fig: Transfer cross-sections} to produce the expected behavior in   dwarfs and clusters, in red and yellow respectively (as discussed in this section).}
    \label{fig: Parameter Space Plot}
\end{figure}

Figure~\ref{fig: Parameter Space Plot} shows the parameter space of the DM particle and $Z'$ sector. The figure depicts the mass values yielding cores in dwarf galaxies at maximum circular velocity of $v=10$ $\mathrm{km/s}$ (red), while avoiding different constraints in galaxy clusters (yellow-shaded region), with $\alpha_{\chi} = 0.01$. In blue, it also shows the required dark matter mass (independent of $Z'$ mass), yielding the right amount of relic abundance from Eq.~\eqref{Relic density today}. Finally, in orange we plot the $(g-2)_{\mu}$ preferred region for a muon coupling constant of $g'_{\mu} = 0.0004$. We observe that the DM coupling constant must be relatively high to generate cross sections as large as those expected from simulations, and the coupling constant to the muon must be four orders of magnitude below, in order to avoid current constraints.

\section{Conclusions}\label{section:Conclusions}

We have explored an atlas of minimal $U(1)'$ extensions to the Standard Model, centered on the anomaly-free parametrization of leptonic flavor symmetries $ L_{i=e, \mu, \tau} $ given by \( \rho(L_{e} - L_{\mu}) + \vartheta(L_{\mu} - L_{\tau}) \). The charges associated with the $U(1)'$ symmetry, mediated by a neutral massive gauge boson \( Z' \), are characterized by the two parameters $\rho$ and $\vartheta$, leading to diverse phenomenological scenarios. Notably, this framework accommodates the possibility of lepton flavor universality violation. These effects have garnered attention due to recent experimental results, including the observed discrepancy in the anomalous magnetic moment of the muon -the $(g-2)_{\mu}$ anomaly-. Within this atlas as a models collection, we aim to address the muon $g-2 $ discrepancy and to mitigate certain challenges in CDM models, particularly regarding small-scale structure formation in galactic halos.

    In the case of the muon anomalous magnetic moment, if the SM calculation based on the data-driven Hadronic Vacuum Polarization contribution is complete, the latest measurements provide an evidence for new physics with a $5.1\sigma$ significance level. Such new physics, as we have shown with the atlas realizations, could be originated from a $Z'$ with a mass on order of $10~\mathrm{MeV}$. However, the product of the charge and coupling of the muon must be around $10^{-4}$ to evade searches from leptophilic mediators. On the electron side for the $(g-2)_e$, the last theory-experiment deviation is set to $2.4\sigma$. We leverage as constraint the fact that the new physics involved in the electron flavor impacting $(g-2)_e$ is not strong enough to spoil the value measured. 
    
    Since the model can accommodate a fermionic dark matter candidate with equal coupling but different charge thereby producing a stabilization mechanism, we add a fermionic self-interacting particle via of the mediator emerging in the gauged $U(1)_l$ extension and discuss some of the cosmological and astrophysical consequences of the complete dark sector.
    
    From a cosmological perspective, the production of the DM particles occurs thermally via a standard freeze-out scenario. In the regime where $m_{Z'} \ll m_{\chi}$, the main annihilation channel is $\chi \overline{\chi} \to Z'Z'$, independent of the $Z'$ mass. The observed value of $\Omega_{DM}h^2$ is then determined solely by an appropriate choice of $g'_{\chi}$ and $m_{\chi}$.
    
    From an astrophysical perspective, the self-scattering of DM fermions mediated by the $Z'$ could influence the formation and distribution of dark matter halos across different scales. Small halos and subhalos initially follow an NFW profile and undergo a two-phase process: core expansion followed by gravitational core collapse. During the core expansion phase, efficient heat transfer redistributes heat from the outer to the inner regions of the halo, reducing central density. Once thermal equilibrium is reached, the second phase begins, during which heat and energy transfer reverses direction, leading to an increase in central density. In contrast, galaxy clusters form through the agglomeration of smaller structures due to gravitational attraction, with thermodynamic effects playing no significant role. The first process requires a large viscosity cross section for self-scattering. The viscosity cross section, a figure of merit describing energy and heat transfer during anisotropic scattering within halos, is preferred over the total cross section. Conversely, galaxy clusters and groups of clusters behave as collisionless fluids, necessitating a negligible self-interaction cross section. The distinct physics observed in these scenarios suggest a velocity-dependent Self-Interacting Dark Matter (vdSIDM) candidate with a small viscosity cross section at high velocities and a large cross section at low velocities.

    The vdSIDM paradigm applies to our model: if the mediator is a light boson of mass $\mathcal{O}(\mathrm{MeV})$ emerging from the leptophilic sector, the scattering cross section becomes velocity-dependent. Through a pair of fitting procedures using data from galaxy clusters, low-surface brightness galaxies, dwarf spheroidals, and Milky Way satellites, we determined the values of the parameters $g'{\chi}$ (or $\alpha{\chi}$), $m_{\chi}$, and $m_{Z'}$ that are consistent with observations across all scales. By fixing $\alpha_{\chi} = 0.01$, we obtained $m_{Z'}\sim 20~\mathrm{MeV}$ for both fits. However, $m_{\chi}\sim 200~\mathrm{GeV}$ was found using an MCMC routine for the cross section per mass in Milky Way satellite (MWs) galaxies, while $m_{\chi}\sim 20~\mathrm{GeV}$ resulted from a $\chi^2$-fit routine for the thermally averaged cross section per mass across galaxy clusters, low-surface brightness galaxies, dwarf spheroidals, and Milky Way satellites. Despite the differences, we found that both values produce sufficiently large cross sections at small scales and adhere to cluster constraints.

    Overall, we found that a single mediator is required in both the fermion sector and the dark matter sector of the leptophilic atlas for the gauged $U(1)'_l$. Furthermore, if the mediator has a mass in the ballpark of $10~\mathrm{MeV}$, it opens a window to simultaneously explain the anomalous $(g-2){\mu}$ experimental results, the production of dark matter through thermalization in the early universe, and CDM tensions at small scales.

\section{Acknowledgments}

We would like to thank Camila Correa for providing the datasets of SIDM cross sections for the satellites of Milky Way galaxies. We are also grateful to Yu-Hai Bo and Manoj Kaplinghat for supplying the datasets with the averaged cross sections of low surface brightness (LSB) galaxies and clusters as a function of velocity. Besides, we are indebted to Diego Milanes and Amalia Betancur for their valuable discussions on flavor anomalies and dark matter deeper challenges. We also thank for the financial support from the internal grant of Faculty of Sciences and CICBA (Antonio Nari\~no University) for the project \emph{Exploraci\'on de F\'isica de part\'iculas: An\'alisis de Datos y Desarrollo Tecnol\'ogico en Colaboraciones Internacionales} with ID number: 2024202.

\bibliographystyle{plainnat} 

\bibliographystyle{apsrev4-1}
\bibliography{main}

\end{document}